\documentclass[aps,prb,twocolumn,footinbib,notitlepage]{revtex4-1}
\usepackage[a4paper,top=1.5cm,bottom=2cm,left=2.15cm,right=2.15cm]{geometry}
\usepackage{graphicx}
\usepackage{amsmath}
\usepackage{amssymb}
\usepackage{verbatim}
\usepackage{color}
\usepackage{mathptmx}
\usepackage{hyperref}
\setcitestyle{super}
\definecolor{darkred}{rgb}{.8,0.1,0.2}

\def\cA{{\mathcal A}}
\def\cL{{\mathcal L}}

\def\cF{{\mathcal F}}

\def\cO{{\mathcal O}}

\def\cP{{\mathcal P}}
\def\cR{{\mathcal R}}
\newcommand{\He}{$^3$He}

\newcommand{\Hea}{$^3$He-A$\,$}
\newcommand{\Hefour}{$^4$He}

\def\vA{{\bf A}}

\def\vM{{\bf M}}
\def\vX{{\bf X}}
\def\vG{{\bf G}}

\def\ve{{\bf e}}
\def\vx{{\bf x}}

\def\vz{{\bf z}}
\def\vk{{\bf k}}

\def\vm{{\bf m}}

\def\vp{{\bf p}}
\def\vr{{\bf r}}
\def\Re#1{{\textrm{Re}}#1}
\def\Im#1{{\textrm{Im}}#1}

\def\point#1#2{{\sf #1}_{\mbox{\tiny #2}}}

\def\parity{{\sf P}} 
\def\time{{\sf T}} 
\def\charge{{\sf C}} 
\def\chiral{{\mathbb{Z}_2}}

\def\text#1{\mbox{\tiny #1}}
\newcommand{\kb}{k_{\text{B}}}
\def\ns{\negthinspace}
\def\nicefrac#1#2{\genfrac{}{}{}{1}{#1}{#2}} 
\def\xiGL{\xi_{\text{GL}}}
\def\btauGL{\bar{\tau}_{\text{GL}}}
\def\bxiGL{\bar{\xi}_{\text{GL}}}
\def\kappaKZM{\kappa_{\text{KZM}}}
\def\chiKZM{\chi_{\text{KZM}}}
\def\deltaKZM{\delta_{\text{KZM}}}
\def\deltaOD{\delta_{\text{OD}}}
\def\deltaUD{\delta_{\text{UD}}}
\newcommand{\tinyonehalf}{\frac{\mbox{\tiny 1}}{\mbox{\tiny 2}}}
\def\betawc{\beta^{\text{wc}}}
\def\betasc{\beta^{\text{sc}}}

\newcommand{\vell}{\pmb{\ell}}
\def\ul#1{\underline{#1}}
\begin{document}
\title{\large Excitation of Collective Modes in a Chiral Superfluid by Thermal Quench}
\author{Noble Gluscevich}
\email{nglusc1@lsu.edu}
\affiliation{Hearne Institute of Theoretical Physics, 
             Louisiana State University,
	     Baton Rouge LA 70803, USA}
\author{J.A. Sauls}
\email{sauls@lsu.edu}
\affiliation{Hearne Institute of Theoretical Physics,
	     Louisiana State University,
	     Baton Rouge LA 70803, USA}
\date{December 1, 2026}
%---------------------------------------------------------------------
\begin{abstract}

Based on time-dependent Ginzburg-Landau field theory we show that rapid cooling through the second-order phase transition into superfluid \Hea\ excites collective modes of newly formed chiral domains, in addition to topological defects that are formed via the Kibble-Zurek mechanism.
Simulations of temperature quenches in the presence of Gaussian space-time white noise generate a highly excited inhomogeneous condensate. 
Large-scale simulations exhibit a complex network of domain walls and vortices.
We report results for the excitation of bosonic collective modes by thermal noise as well as nonequilibrium temperature quenches, followed by coarsening dynamics tracked in terms of the Fourier components of the order parameter amplitudes.
For thermal states, the spectrum of bosonic excitations is defined by a power spectral density (PSD) for each mode, which is sensitive to the Langevin damping. 
For weak damping the PSD onsets sharply at the frequency corresponding to the mass of the bosonic mode, then decays as $1/\omega$.
We also track the dynamics of the order parameter following a temperature quench. 
We report results for the scaling exponents of the Kibble-Zurek freeze-out time and correlation length as a function of quench rate for several damping rates.
The dynamic exponent $z$ is shown to transition smoothly from $z=1$ to $z=2$ as the damping is increased, while the correlation length exponent, $\nu\approx 1/2$, is independent of damping.
\end{abstract} 
%---------------------------------------------------------------------
\maketitle

\section{Introduction}
\vspace*{-3mm}

Spontaneous symmetry breaking is central to physical phenomena ranging from mass generation in particle physics and cosmology~\cite{vilenkin94} to phase transitions and emergent quanta in quantum fluids and solids.~\cite{vollhardt90,hin24}
On approaching the onset of a phase transition, thermal fluctuations of the broken symmetry phase become long lived. 
If the system cools through the transition at finite rate, long-lived fluctuations form nuclei for the growth of an inhomogeneous broken symmetry phase hosting a finite density of topological defects characteristic of the otherwise uniform broken symmetry vacuum state.
This idea was put forward by Kibble as a mechanism for the formation of cosmic strings in field theory models of the early universe.~\cite{kib76,kib80}
Zurek made the connection between Kibble's mechanism for cosmic string generation and the generation of quantized vortices in superfluid \Hefour, as well as other condensed matter systems hosting broken symmetry phases.~\cite{zur85}
Indeed the Kibble-Zurek mechanism (KZM) for topological defect generation has been observed in a wide variety of systems, including superfluid $^3$He,~\cite{bau96,ruu96,ruu98,aut16,aut21} cold atomic gases,~\cite{wei08,lam13,lee24} and nematic liquid crystals.~\cite{chu91,bow94}
Furthermore, the inhomogeneous condensate that is created following a quench is a mechanism for the generation 
of quantum turbulence,~\cite{shi26,sau26} as well as the excitation of bosonic 
collective modes, which is a focus of this report.

The chiral A-phase of superfluid \He\ is the equilibrium state over the entire pressure and temperature phase diagram when confined to films of thickness $d < D_{c_2} \approx 9\xi_0(p)$, where $\xi_0=\hbar\,v_f/2\pi\kb T_c$ is the pressure dependent pair correlation length at $T=0$. 
Thus, at $p\simeq 0\,\mbox{bar}$, $D_{c_2}\approx 0.7\,\mu\textrm{m}$.~\cite{vor07} 
Furthermore, specular boundaries are not pair-breaking so the transition temperature is given by the bulk value, $T_c\simeq 1\,\mbox{mK}$.~\cite{hei25}
In zero magnetic field the ground state manifold of superfluid \He\ is restricted due to the nuclear dipolar energy to an equal-spin pairing (ESP) state with zero spin projection along the normal to the film surface, and to the two-dimensional (2D) p-wave basis, $\Delta(\vp)=A_x\,\vp_x + A_y\,\vp_y$. 
The resulting order parameter is then defined by a 2D complex vector, 
\begin{equation}
\vA = A_x\,\ve_x + A_y\,\ve_y
\,,
\label{eq-film-order_parameter}
\end{equation}
where $\ve_x$ and $\ve_y$ are fixed orthonormal Cartesian unit vectors in the plane of the \He\ film. 
It is well established theoretically that strong-coupling corrections to weak-coupling BCS theory for \He\ films favor the 2D chiral A-phase,
$\vA=\Delta\,\left(\ve_x + i \ve_y\right)/\sqrt{2}$,
as the equilibrium state for films of thickness $d \le D_{c_2}(p,T)$.~\cite{wim16}
This identification of the ground state of \He\ films is confirmed by NMR experiments.~\cite{hei25}

In addition to breaking the global $\point{U(1)}{\ns}$ symmetry, the chiral phase breaks spin and orbital rotation symmetries, as well as parity ($\parity$) and time-reversal ($\time$) symmetry. The residual symmetry group of the 2D chiral phase includes spin rotations about the normal to the film, $\point{U(1)}{\ns S\mbox{$_z$}}$, joint gauge-orbit rotations, $\point{U(1)}{\ns N-L\mbox{$_z$}}$, and a $\chiral$ symmetry corresponding to the product of time-reversal and mirror reflection in a plane containing the $z$-axis. 
The breaking of time-reversal symmetry implies the ground state is doubly degenerate,
\begin{equation}
\vA_{\pm}=\Delta\,\ve_{\pm}
\quad\mbox{with}\quad
\ve_{\pm}\equiv\left(\ve_x\pm i\ve_y\right)/\sqrt{2}
\,,
\end{equation}
corresponding to condensates of Cooper pairs with opposite (time-reversed) internal orbital angular momentum, 
$L_z\,\ve_{\pm} = \pm\hbar\,\ve_{\pm}$.~\footnote{We set $\hbar=1$ in the main text unless noted otherwise.}
Thus, in addition to phase vortices classified by $\pi_1(S^1)=\mathbb{Z}$, \Hea\ films support topologically stable \emph{domain walls} separating time-reversed chiral ground states.

Recently we reported numerical simulations of thermal quenches of \Hea\ films based on time-dependent Ginzburg-Landau (TDGL) theory that provided new results for
Kibble-Zurek (KZ) scaling and full counting statistics of the formation of topological defects, vortices, anti-vortices and domain walls, as well as their post freeze-out dynamics for a range of quench rates.~\cite{glu25} 

We report new results showing that collective modes of an inhomogeneous chiral order parameter are excited by a temperature quench immediately following KZ freeze-out. 
The temperature quench generates a complex spectrum of excitations governed by stochastic TDGL field equations coupled to a Langevin noise source. 
We begin with a brief summary of the TDGL theory and the spectrum of bosonic modes of an equilibrium chiral superfluid.  

%\vspace*{-7mm}
\section{TDGL Theory for $^3\mbox{He}$ Films}
\vspace*{-3mm}

For thin \He\ films in zero-field the general form of the order parameter is a two-dimensional complex vector 
defined in Eq.~\eqref{eq-film-order_parameter}. 
The components, $A_{i}$, transform as a vector under rotations about the axis normal to the plane of the film, $A_i\xrightarrow[]{\phi}\cR[\phi\ve_z]_{ij}\,A_j$, 
while under gauge transformations $A_i\xrightarrow[]{\chi}e^{i\chi}\,A_i$. 
These symmetries constrain the terms contributing to the bosonic action for space-time variations of the order parameter, $S=\int dt\,d^3r\,\cL[\vA,\partial_t\vA,\partial_i\vA]$, where the Lagrangian density is determined by the linearly independent 2$^{nd}$- and 4$^{th}$-order invariants constructed from $\vA$, $\vA^*$ and their space-time derivatives,~\cite{glu25,sau15}  
\begin{eqnarray}
\mathcal{L} 
&=&
\mu\,(\partial_t A_i)(\partial_t A_i^*)
-
K_1\,(\partial_i A_j)(\partial_i A_j^*)
\nonumber\\
&-&
K_2\,(\partial_i A_i)(\partial_j A_j^*)
-
K_3\,(\partial_i A_j)(\partial_j A_i^*)
\nonumber\\
&-&
\alpha\,A_i\,A_i^*
-
\bar\beta_{1}(A_i A_i^*)^2
-
\bar\beta_{2}(A_i A_i)(A_j^* A_j^*)
\,.
\label{eq-Lagrangian}
\end{eqnarray}
The material coefficients are calculated from BCS theory for spin-triplet, p-wave pairing 
in \He~\cite{wim16,sau17,reg20,wim19} and summarized in App.~\ref{app-TDGL_parameters}.
Note that $\bar\beta_1 \equiv \beta_{245}$ and $\bar\beta_2 \equiv \beta_{13}$ are the relevant combinations 
of fourth-order coefficients of bulk \He\ that define the TDGL action for strongly confined \He\ films. 
We also use $K_{23} \equiv K_2 + K_3$.

\vspace*{-5mm}
\subsection{Order parameter collective modes}\label{sec-OP_modes}
\vspace*{-3mm}

We first consider small space-time fluctuations of the order parameter in the presence of a homogeneous equilibrium chiral ground state with orbital angular momentum $L_z=+\hbar$ per pair,
\begin{equation}
\vA(\vr,t)=\vA_{+}+\cA(\vr,t)
\,,
\end{equation}
where $\vA_+=\Delta\ve_{+}$.
The space-time fluctuations are encoded in two complex amplitudes which we resolve in the chiral basis,
\begin{equation}\label{eq-OP_fluctuations}
\cA(\vr,t) = D(\vr,t)\,\ve_{+} + E(\vr,t)\,\ve_{-}
\,.
\end{equation}
Thus, there are two classes of excitations: fluctuations with the same chirality as the ground state, $\ell_z=+1$, represented by the field $D(\vr,t)$, and fluctuations with time-reversed chirality, $\ell_z=-1$, represented by the field $E(\vr,t)$. 
Altogether there are four collective modes corresponding to the real and imaginary components of the two complex amplitudes: 
$D^{+}=(D+D^*)/2$, 
$D^{-}=(D-D^*)/2i$, 
$E^{+}=(E+E^*)/2$,
$E^{-}=(E-E^*)/2i$.
These amplitudes have definite parity under charge conjugation, i.e. particle-hole 
transformation: $\charge\,D^{\pm}=\pm\,D^{\pm}$, and similarly for the $E$ modes.~\cite{ser83a,fis85}
Thus, we can label the four fluctuation amplitudes by the pair of discrete quantum numbers: chirality, $\ell_z=\pm 1$, and charge conjugation parity, $c=\pm 1$.
In the long-wavelength limit, the gradient terms coupling $D^{\pm}$ and $E^{\pm}$, which are proportional to $K_{23}$, can be neglected. The resulting Lagrangian from Eq.~\eqref{eq-Lagrangian} expanded to leading order reduces to\footnote{A detailed discussion of the gradient generated mode mixing terms is provided in App.~\ref{app-K23}.} 
\begin{eqnarray}
L=\int_{\text{V}}d^3r\,
\Big\{
\mu\,\Big[(\dot{D}^{+})^2 + (\dot{D}^{-})^2 
+(\dot{E}^{+})^2 + (\dot{E}^{-})^2\Big]
\nonumber\\
-4\Delta^2\Big[\bar\beta_1(D^{+})^2 +\bar\beta_2\left((E^{+})^2 + (E^{-})^2
\right)\Big] 
\nonumber\\
-2 K_1\Big[(\partial_i D^+)^2 + (\partial_i D^-)^2 
         + (\partial_i E^+)^2 + (\partial_i E^-)^2
      \Big]
\Big\} 
\,.
\label{eq-Lagrangian_modes}
\end{eqnarray}

Hereafter we consider \He\ films with thicknesses sufficiently small that we can neglect spatial variations of the order parameter in the direction normal to the film, and replace $d^3r = d\times d^2r$ where $d$ is the film thickness, which is still large compared to atomic scales.

In the homogeneous limit the Euler-Lagrange equations then reduce to four uncoupled mode equations,
\begin{equation}
\ddot{D}^{c} + M^2_{\text{D}^{c}}\,D^{c} = 0
\quad\mbox{and}\quad
\ddot{E}^{c} + M^2_{\text{E}^{c}}\,E^{c} = 0
\,,
\end{equation}
where $M_{\text{D}^{c}}$ and $M_{\text{E}^{c}}$ are the excitation gaps (``masses'') of the bosonic modes with quantum numbers $\ell_z=\pm 1$ and $c=\pm 1$ as summarized in Table~\ref{table-Mode_Spectrum}.
Note that $D^{-}$, which is a cyclic coordinate of the Lagrangian in Eq.~\eqref{eq-Lagrangian_modes}, is the Nambu-Goldstone mode associated with the broken $\point{U(1)}{\ns}$ symmetry.
For small amplitude, $|h(\vr,t)|\ll 1$ and phase, $|\theta(\vr,t)|\ll 1$, fluctuations of the $\ell_z=+1$ ground state, $\vA(\vr,t)=\Delta(\vr,t)\,e^{i\theta(\vr,t)}\,\ve_{+}$ $\approx \Delta\,\ve_{+} \left(1+h(\vr,t)+i\theta(\vr,t)\right)$. 
Thus, $D^{-}=\Delta\,\theta(\vr,t)$ is the phase fluctuation, and $D^{+}=\Delta\,h(\vr,t)$ is the amplitude fluctuation. 
The amplitude mode is the Higgs boson for this chiral ground state. In particular, $D^{+}$ is the amplitude of an excitation with the same quantum numbers ($\ell_z=+1$ and $c=+1$) as those of the ground state.

%---------------------------------------------------------------------
\begin{table}[t]
\begin{center}
\begin{tabular}{c|c|c|c}
Mode & Symmetry	& Mass & Name
\\
\hline
$D^{+}$	& $\ell_z=+1$ $c=+1$ & $2\Delta$ & Higgs Mode
\\
$D^{-}$	& $\ell_z=+1$ $c=-1$ & $0$	& NG Mode
\\
$E^{+}$	& $\ell_z=-1$ $c=+1$ & $\sqrt{2}\Delta$ & E$^{+}$ Mode
\\
$E^{-}$	& $\ell_z=-1$ $c=-1$ & $\sqrt{2}\Delta$ & E$^{-}$ Mode
\\
\hline
\end{tabular}
\end{center}
\caption{Bosonic Mode Spectrum for the $\vA_{+}=\Delta\,\ve_{+}$ chiral ground state.
The Higgs mode with $M_{\text{D}^{+}}=2\Delta$ has the same quantum numbers as the chiral ground state.
The masses of the $E^{\pm}$ modes are degenerate and given by $M_{\text{E}^{\pm}}=\sqrt{2}\Delta$ 
in the weak-coupling limit for \He.
}
\label{table-Mode_Spectrum}
\end{table}
%---------------------------------------------------------------------

In the weak-coupling limit the mass of the Higgs mode is $M_{\text{D}^{\text{+}}}=2\Delta$, where $\Delta$ is the gap (mass) in the fermionic spectrum of the broken symmetry phase.~\cite{hig64,lit81} 
Since the Higgs mode has the same quantum numbers as the condensate, the Higgs mass is unshifted by polarization effects of the underlying fermionic vacuum.~\cite{sau15}
Thus, the inertia term in the effective Lagrangian is given by $\mu=\bar\beta_1$, and sets the mass scale for all the modes of the effective Lagrangian, including the modes with opposite chirality, $\ell_z=-1$. 
The $E$-modes are the clapping modes of an isotropic 2D chiral p-wave ground state with masses given by~\footnote{{The label ``clapping modes'' originates from a geometric interpretation of the orbital motion by P. W\"olfle~\cite{wol77}. The clapping modes, and all gapped bosonic modes of superfluid \He, are broadly referred to as Higgs modes differentiated from one another by their quantum numbers defined by the residual symmetry of the ground state, see Refs.~\onlinecite{vol13,sau17}. The original Higgs mode is the gapped excitation with the same quantum numbers as the ground state, which in the case of the $\ell_z=+1,c=+1$ ground state is the $D^{+}$ mode.}}
\begin{equation}
M_{\text{E}^{\pm}}=2\sqrt{\bar\beta_2/\bar\beta_1}\,\Delta\,\rightarrow\sqrt{2}\Delta
\,.
\end{equation}
The limiting value of $\sqrt{2}\Delta$ is the prediction of weak-coupling BCS theory for an isotropic 2D chiral p-wave ground state in which $\bar\beta_2/\bar\beta_1=\nicefrac{1}{2}$.
Thus, the two time-reversed modes are \emph{degenerate} and lie well below the pair-breaking continuum edge at $2\Delta$.~\footnote{In anisotropic chiral superconductors the degeneracy of the $E^{\pm}$ modes is lifted by anisotropy of the Fermi surface and pairing basis functions as discussed in Ref.~\onlinecite{sau15}.}

For small amplitude fluctuations of the $\vA_+$ ground state the gradient terms in Eq.~\eqref{eq-Lagrangian} give rise to well defined energy-momentum dispersion relations for the bosonic modes, 
\begin{equation}
E_{\mu}(\vk)=\sqrt{M_{\mu}^2 + v_{\mu}^2\,|\vk|^2}\,,\quad\mu\in\{D^{\pm},E^{\pm}\}
\,,
\end{equation}
with velocities $v_\mu = \sqrt{2K_1/ \bar{\beta}_1}=v_f$ for all four modes.
Under nonequilibrium conditions nonlinear terms in the expansion of the Lagrangian density in terms of space-time fluctuations of the mode amplitudes can become important, leading to mode-mode coupling. 
Nonlinear effects on the excitation spectrum are discussed in Sec.~\ref{sec-nonlinear}.

\vspace*{-5mm}
\subsection{Langevin dynamics}
\vspace*{-3mm}

The TDGL equations describe the conservative dynamics of the order parameter for superfluid \He. 
However, the order parameter nucleates and grows in the presence of thermal noise. 
We describe the dynamics of the order parameter in the presence of thermal noise via a stochastic Langevin source term, $\zeta_i(\vr,t)$, that drives random fluctuations of the order parameter. 
The strength of the Langevin noise source also determines the dissipative response to large fluctuations.~\footnote{The temperature and pressure dependence of the noise and damping rate have not been systematically investigated. We expect $\Gamma$ to decrease with temperature below $T_c$ and to increase with increasing pressure.}
The resulting stochastic TDGL equations, 
\begin{eqnarray}
\mu\ddot{A}_i 
+ \Gamma\dot{A}_i + \alpha(T) A_i
- K_1\partial^2 A_i - K_{23}\,\partial_i(\partial_j A_j)
\qquad\qquad
\\
+2\bar\beta_{1}(A_j A_j^*)\,A_i 
+2\bar\beta_{2}(A_j A_j)\,A_i^*
= 
\zeta_i(\vr, t)
\,,i\in\{x,y\}
\,,
\nonumber
\label{eq-TDGL_films}
\end{eqnarray}
include both oscillatory and diffusive dynamics related to second-order ($\mu\ddot{A}_i$) and first-order time derivatives ($\Gamma\dot{A}_i$), respectively.

In numerical solutions of the stochastic field equations the damping rate plays a central role in determining scaling exponents of the system.~\cite{suz25} 
In our analysis we consider the dynamics of the order parameter for a range of values for the noise strength and damping rate, $\Gamma$. 
The latter can be expressed as a dimensionless damping parameter, $\gamma$, by scaling $\Gamma$ by the microscopic timescale for order parameter growth, i.e. the pair formation timescale, $\btauGL$, defined in Eq.~\eqref{eq-tauGL}, and the coefficient determining the inertial response of the order parameter,
\begin{equation}
\gamma=\Gamma\,\btauGL/\mu
\,.
\end{equation}

In the following analysis we vary $\gamma$ from the strong damping limit, $\gamma=1.0$, to the weak damping limit, $\gamma=0.005$. 
The order parameter is coupled to an uncorrelated Gaussian space-time white noise source, $\zeta_i(\vr,t)$, with variance determined by the damping coefficient $\Gamma$ and $\kb T$,
\begin{equation}
\langle\zeta_i(\vr,t)\,\zeta^*_j(\vr',t')\rangle 
=
2\Gamma\,\kb T\,\delta_{ij}\,\delta(\vr-\vr')\,\delta(t-t')\
\,.
\end{equation}

The noise correlator can also be expressed in terms of spatial Fourier coefficients,
\begin{equation}
\hspace*{-2.75mm}
\langle\zeta_i(\vk,t)\zeta_j^*(\vk',t')\rangle 
\ns\ns
=
\ns 
2\Gamma\kb T\delta_{ij}(2\pi)^2\delta(\vk-\vk')\delta(t\ns-\ns t')
\end{equation}
which is useful for the pseudospectral methods which we use in this work. 

\vspace*{-5mm}
\subsection{Simulation methods}
\vspace*{-3mm}

In the thin film limit of \He\ considered here the stochastic TDGL equations are integrated with respect to time and two spatial dimensions, the latter with periodic boundary conditions.
Integration in time is performed with a stochastic second-order Verlet integrator which is symplectic in the limit of zero damping.~\cite{bus07}
We use the time step $\Delta t = 0.02 \btauGL$, except for homogeneous simulations in Sec. \ref{sec-nonlinear} where we use $\Delta t = 10^{-4} \btauGL$.
For spatial variations we Fourier transform the order parameter field, 
\begin{equation}
\cF[\vA](\vk,t)=\int d^2r\,e^{-i\vk\cdot\vr}\,\vA(\vr,t)
\equiv 
\vA(\vk,t) 
\,,
\end{equation}
and use Fourier pseudospectral methods to solve the stochastic TDGL equations,~\cite{ors69,got77} 
\begin{eqnarray}
\mu\partial_t^2 A_{i}(\vk,t) + \Gamma\partial_t A_{i}(\vk,t) + \alpha(T) A_{i}(\vk,t)
\hspace*{1.6cm}
\nonumber\\
+K_1 (k_x^2 + k_y^2) A_{i}(\vk,t) + K_{23} k_i k_j A_{j}(\vk,t) 
\hspace*{1.6cm}
\nonumber
\\
+\cF\left[2 \bar\beta_{1} A_j A_j^* A_i + 2 \bar\beta_{2} A_j A_j A_i^*\right](\vk,t)
=\zeta_{i}(\vk,t)
\,.\,\,
\label{eq-stochastic_TDGL}
\end{eqnarray}
The Fourier coefficients of the nonlinear terms must be computed in real space and Fourier transformed back, where de-aliasing is handled with explicit zero-padding.~\cite{got77}
The advantage of the pseudospectral method with periodic boundary conditions is that the basis functions are Fourier modes, providing spectral properties at no extra computational cost.
For the calculation of equilibrium properties and the quench dynamics of the $\vk=0$ mode at the Kibble-Zurek freeze-out time, we use a system size $L\times L=128 \bxiGL \times 128 \bxiGL$, with $N_k = 256$ grid points in each dimension, i.e. a uniform grid spacing of $h=0.5\bxiGL$.
When solving for the generation of modes out of equilibrium and spatial correlation functions, a larger system size is required, where we use $L \times L = 440\,\bxiGL \times 440\,\bxiGL$, with $N_k=880$ grid points along each direction. 
All results reported here are based on averaging over an ensemble $\Omega$ of at least 20 simulations, unless stated otherwise.
Ensemble averages are denoted by $\langle \dots \rangle_\Omega$.

For excitation of the bosonic modes $D^\pm$, $E^\pm$ by thermal noise we solve the full stochastic TDGL equations and define $D^+$ ($D^-$) to be the amplitude (phase) fluctuations of $A_+$, while $E^+$ ($E^-$) are the real (imaginary) parts of $A_-$.
These four amplitudes are evaluated in real space, then Fourier transformed to obtain the Fourier coefficients. 
The system is initialized with a uniform order parameter at its equilibrium value, then allowed to thermalize for a time of $500\,\btauGL$.
Over an additional time period of $2000\,\btauGL$ the dynamics of the modes are calculated, and the power spectral density (PSD) is computed by taking the Fourier transform in time with a Hann window function~\cite{press86} and integrating over all wavevectors.

When performing a temperature quench, the system is initialized at $T=1.5T_c$, and allowed to thermalize for $100 \bar{\tau}_{\textrm{GL}}$.
Then the temperature is cooled linearly in time to a final temperature of $T = 0.5T_c$ over a time period of $\tau_Q$ with $T(t=0)\equiv T_c$,
\begin{equation}
\epsilon(t) \equiv \frac{T(t)}{T_c} - 1 = -\frac{t}{\tau_Q}
\,,\mbox{for}\,-\tau_Q/2 \leq t \leq \tau_Q/2
\,.
\end{equation}

%---------------------------------------------------------------------
\vspace*{-5mm}
\section{Excitation of Modes by Thermal Noise}\label{sec-modes_equilibrium}
\vspace*{-3mm}

\begin{figure}
\centering
\includegraphics[width=\linewidth]{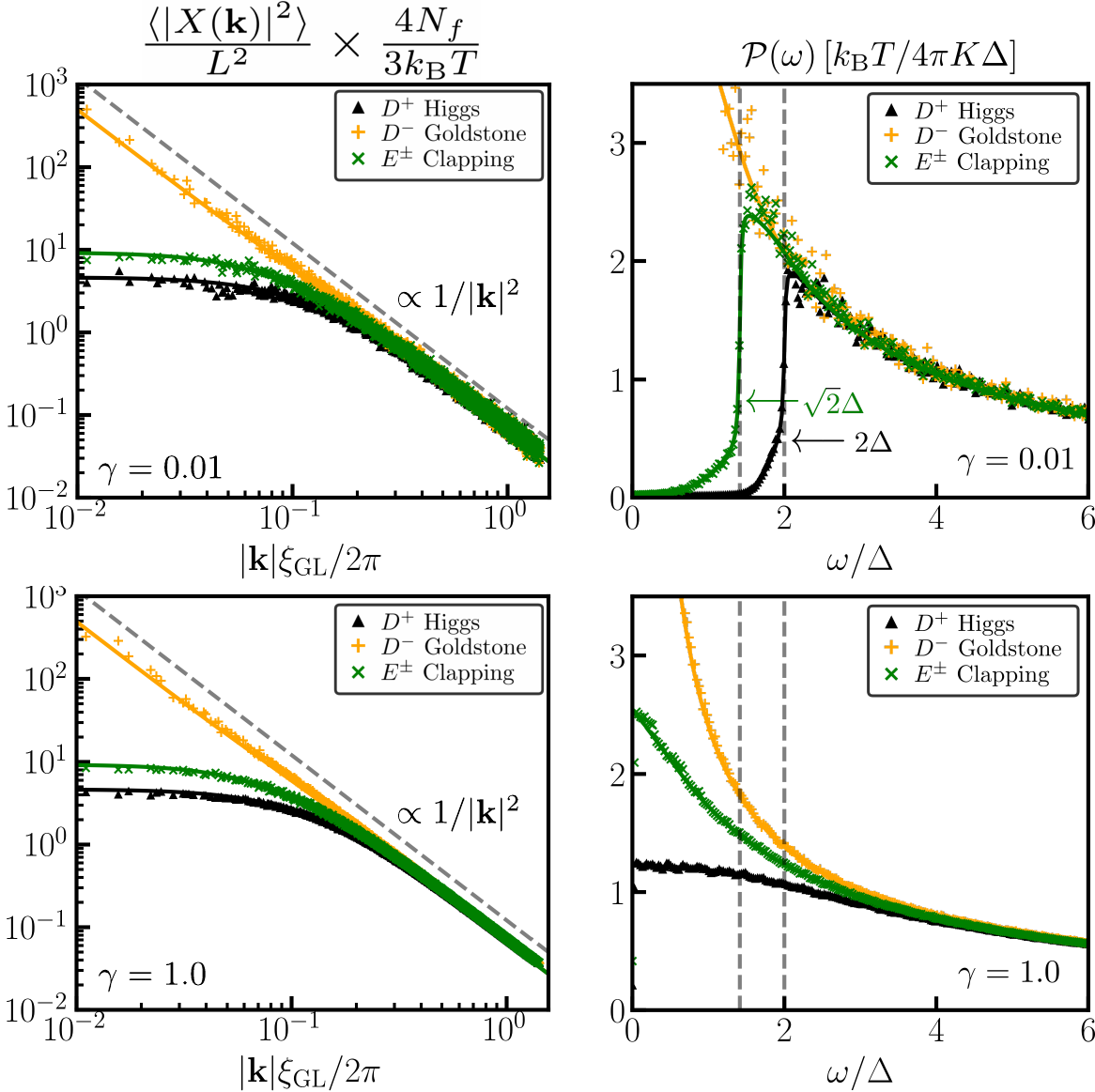}%equil_with_better_analytics.png}
\caption{
    Upper row, left panel: thermal populations of the collective modes [$X=D^\pm,E^\pm$] for the weak damping limit, $\gamma = 0.01$.  The mode populations decay $\propto 1/k^2$ in the short wavelength limit, $|\vk|\gg M_X/v$, as shown by the $\propto 1/k^2$ dashed grey line as a guide to the eye.
Upper row, right panel: The masses of the Higgs ($D^{+}$) and clapping modes ($E^{\pm}$) exhibit sharp thresholds in the power spectrum, followed by decay $\propto 1/\omega$ at higher frequencies.  The PSD for the massless Goldstone mode ($D^{-}$) scales as $\kb T/\omega$.
Bottom row: The same for the modes in the overdamped regime with $\gamma=1.0$.  The thermal population of modes is independent of damping, but the sharp mass threshold in weak damping is replaced by a smooth crossover for strong damping.
The solid curves are the analytic results discussed in App.~\ref{app-K23}.
}
\label{fig-equilibrium_mode_PSD}
\end{figure}

The autocorrelation function, $\langle X(\vk, t) X^*({\vk', t')} \rangle$, of the bosonic excitation
$X(\vk,t)$, where $X\in\{D^{+}, D^{-}, E^{+}, E^{-}\}$, is related by Fourier transform to the PSD of the bosonic modes in thermal equilibrium in the frequency domain.
For an excitation $X(\vr,t)$, the average power associated with the signal is
\begin{equation}
E_X=\lim\limits_{L\rightarrow \infty} \lim\limits_{\tau \rightarrow \infty}
    \frac{N_f}{L^2 \tau}\int d^2r \int dt
    \left|X(\vr,t)\right|^2
\,,
\end{equation}
where $N_f$ is the density of states at the Fermi energy.
We also focus on calculating the PSD in the frequency domain integrated over all wavevectors,
\begin{equation}
\mathcal{P}_X(\omega)
=\lim\limits_{L \rightarrow \infty} \lim\limits_{\tau \rightarrow \infty} 
\frac{N_f}{L^2 \tau}
\int\dfrac{d^2k}{(2\pi)^2} \left| X(\vk,\omega) \right|^2 
\,.
\label{eq-PSD_k-space}
\end{equation}
Hereafter we set the prefactor $N_f=1$ in the PSD and mode population.
Results for the equilibrium correlation function and PSD for all bosonic modes can be obtained analytically if the nonlinear terms in the stochastic TDGL equations are negligible.~\cite{kardar07}
Here we provide analytic results neglecting the gradient generated mode-mixing terms that capture the full structure of the PSD and equilibrium populations for the modes. 
Exact analytical results that include gradient-induced mode-mixing are derived in App.~\ref{app-K23}. Eqs.~\eqref{eq-PSD_vs_s-omega},~\eqref{eq-PSD_final} and~\eqref{eq-thermal_population_X} are used to generate the analytic results plotted in Fig.~\ref{fig-equilibrium_mode_PSD}.
The linearized equation of motion for an excitation in Fourier coefficients is
\begin{equation}
\hspace*{-2mm}
\mu\ddot{X}(\vk)\ns +\ns \Gamma \dot{X}(\vk)\ns +\ns \mu M_X^2 X(\vk)\ns 
+\ns K |\vk|^2 X(\vk)\ns=\ns\zeta_{X} (\vk, t)
\,,
\end{equation}
where $K=2K_1$. The excitation is described by a damped harmonic oscillator driven by a white noise term, which has power spectral density.~\cite{wan45}

%---------------------------------------------------------------------
%---------------------------------------------------------------------

\begin{equation}
\frac{\langle |X(\vk,\omega)|^2 \rangle}{L^2} 
= 
\frac{\Gamma\,\kb T}{(\mu\omega^2 -  \mu M_X^2 - K |\vk|^2)^2 + \Gamma^2 \omega^2}
\,,
\end{equation}
with $L^2 \rightarrow (2\pi)^2 \delta(\vk-\vk')$ in the continuum limit. 
The thermal population of a mode is obtained from the static limit of the correlation function by integrating the PSD over all frequencies,~\footnote{This result also follows from the Boltzmann distribution based on the TDGL Hamiltonian, as well as the equipartition theorem.}$^,$~\cite{kardar07,wan45}
\begin{equation}
\frac{\left\langle |X(\vk)|^2\right\rangle}{L^2}
=
\frac{\frac{1}{2}\kb T}{\mu M_X^2 + K|\vk|^2}
\,.
\end{equation}
Thus, the thermal noise-induced population of the massless Goldstone mode is proportional to $\kb T$ and scales as $\propto 1/|\vk|^2$.
Numerical solutions of the stochastic TDGL equation~\eqref{eq-stochastic_TDGL} used to calculate the population density of the Nambu-Goldstone mode ($D^{-}$) exhibit this scaling behavior in both the overdamped regime ($\gamma = 1$) and the weakly damped regime ($\gamma = 0.01$), as shown in the left panels of Fig.~\ref{fig-equilibrium_mode_PSD}.
In contrast, the noise-induced populations of the massive Higgs and clapping modes saturate in the long wavelength limit $K|\vk|^2 \ll \mu M_X^2$, but ultimately decay $\propto 1/|\vk|^2$ in the short wavelength limit.
Integrating the momentum-dependent PSD over all $\vk$ and dividing by $(2\pi)^2$ yields the total PSD for the bosonic excitation,
\begin{equation}
\mathcal{P}_{X}(\omega)
=
\frac{\kb T}{4\pi K\omega}
\left[\frac{\pi}{2}-\tan^{-1}\left(\mu\frac{M_X^2-\omega^2}{\Gamma\omega}\right)\right]
\,.
\label{eq-total_PSD_Gaussian} 
\end{equation}
This result exhibits a spectral signature of the bosonic modes. 
In particular, for weak damping, $\Gamma/\mu\Delta\ll 1$, and $\omega\lesssim M_X$ the arctangent approaches $\pi/2$ leading to near zero power below the mass threshold at $\omega=M_X$.
For frequencies immediately above the mass threshold the arctangent switches sign generating a sharp increase in power at $\omega=M_X$, followed by decay as $\sim 1/\omega$ well above the mass threshold.
These features are clearly observed in the PSD obtained from numerical solutions to the stochastic TDGL Eqs.~\eqref{eq-stochastic_TDGL}. 
In particular, both the Higgs mode with mass $M_{\text{D}^+} = 2\Delta$ and the pair of clapping modes with $M_{\text{E}^{\pm}} = \sqrt{2}\Delta$ show the sharp thresholds in the PSD at their respective masses in the upper right panel of Fig.~\ref{fig-equilibrium_mode_PSD}, followed by decay $\propto 1/\omega$ at higher frequencies.
Note that mode-mixing (c.f. App.~\ref{app-K23}) increases the spectral weight, both above and below the mass threshold of the Higgs and clapping modes, even in the limit of low damping.
As expected there is no threshold in the power spectrum for the massless phase mode.
In the overdamped regime the mass thresholds for the Higgs and clapping modes are washed out, replaced by a smooth cross-over to a limiting value scaling as 
\begin{equation}
\lim_{\omega\rightarrow 0}\cP_X(\omega)
=
\frac{\Gamma \kb T}{4\pi \mu K M_X^2}\,,\quad X\in\{D^{+},E^{\pm}\}
\,.
\label{eq-total_PSD_Gaussian_zero_frequency} 
\end{equation}
Note the increase in the limiting power of the clapping mode compared to the Higgs mode by a factor very close to 2, which is what is expected from the mass dependence in Eq.~\eqref{eq-total_PSD_Gaussian_zero_frequency}. 

%---------------------------------------------------------------------
\begin{figure}
\centering
\includegraphics[width=\columnwidth]{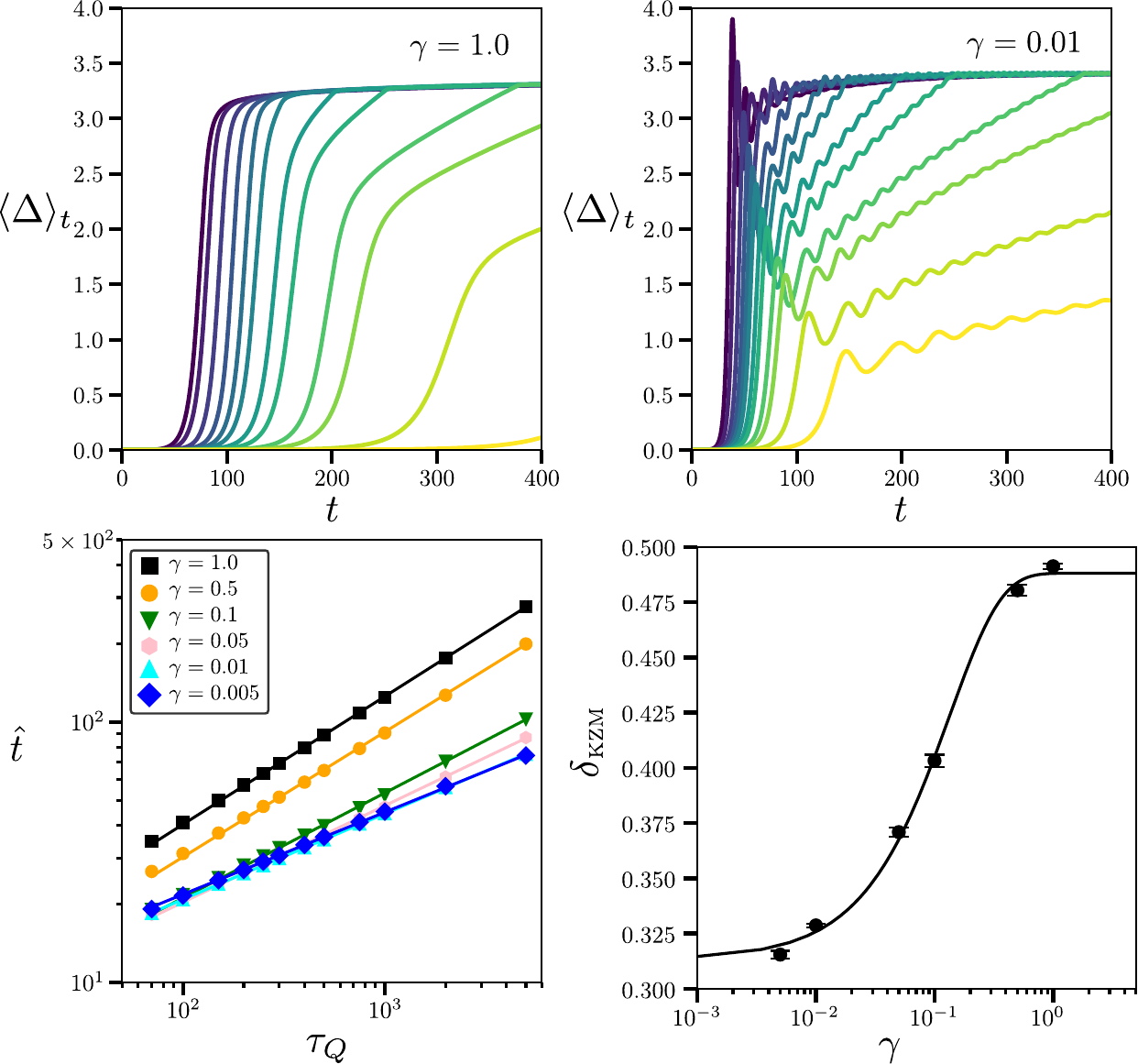}
\caption{Upper row: the spatially averaged order parameter, $\langle\Delta\rangle_t$ (in units $\kb T_c$) as a function of time (in units $\btauGL$) after passing through $T_c$, for quench times:
$\tau_Q/\btauGL\in\{70, 100, 150, 200,$ $250, 300, 400, 500, 750, 1000, 2000, 5000\}$.
The quench rate is indicated by color, varying from purple (fastest) to yellow (slowest).
The left panel is for high damping ($\gamma=1.0$) and the right for low damping ($\gamma=0.01$).
The oscillations apparent in the underdamped regime are excitations of the Higgs and clapping modes. More detailed analysis of the oscillations is provided in Sec.~\ref{sec-modes_via_quench} and shown in Fig.~\ref{fig-onset_of_modes}.
Lower row: The KZ freeze-out time obtained from $\langle \Delta \rangle = 0.01 \kb T_c$ scales as $\tau_Q^{\deltaKZM}$ with an exponent $\deltaKZM$ that depends on the damping parameter (left panel).
The right panel shows the cross-over in the exponent from $\deltaKZM=0.488$ in the overdamped limit to $\deltaKZM=0.313$ in the underdamped limit, both in reasonable agreement with the mean-field predictions of $1/2$ and $1/3$ in these limits.~\cite{suz25}
}
\label{fig-freeze-out}
\end{figure}
%---------------------------------------------------------------------

%---------------------------------------------------------------------
\begin{figure*}[t!]
\centering
\includegraphics[width=\textwidth]{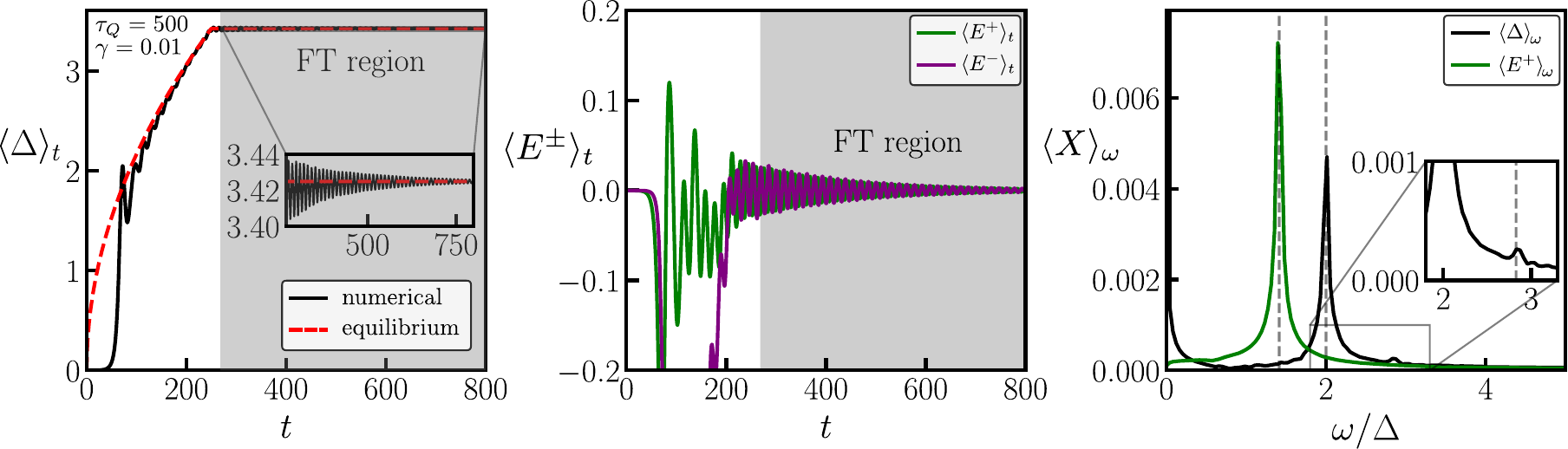}
\caption{
Onset of collective modes for a single quench simulation with 
$\tau_Q=500\,\btauGL$ and $\gamma = 0.01$.
Left panel: 
The black solid line is the simulation result for $\langle\Delta\rangle_t$ compared with the equilibrium value at temperature $T(t)$ shown by the red dashed line.
Inset: highlight of the oscillations of $\langle\Delta\rangle_t$ starting shortly after the target temperature at $t = \tau_Q/2 = 250\,\btauGL$ is reached. The time window $t\in [270,800]\btauGL$ is used for Fourier analysis of the oscillations.
Middle panel: Spatially averaged real ($E^+$) and imaginary ($E^-$) components of $A_-$, the subdominant order parameter component in the dominant $A_+$ domain.
At early times $A_+$ and $A_-$ domains compete, but in this simulation the $A_+$ domain survives, leaving large fluctuations of the $A_-$ amplitude.
Right panel: Fourier transforms of $\langle\Delta\rangle_t$ (black) and $\langle E^+ \rangle_t$ (green).
The clapping modes also appear in $\langle\Delta\rangle_\omega$ at the smaller resonant peak at $\omega = 2\sqrt{2}\Delta$ originating from the $(E^{\pm})^2$ contributions to $\Delta(\vr,t)$ (see text).
}
\label{fig-onset_of_modes}
\end{figure*}
%---------------------------------------------------------------------

%---------------------------------------------------------------------
\vspace*{-7mm}
\section{Excitation by a Temperature Quench}
\vspace*{-3mm}

In a temperature quench, the spatial average of the order parameter,
\begin{eqnarray}
\langle\Delta\rangle_t &\equiv& \frac{1}{L^2}\int d^2r\,\sqrt{\vA(\vr,t)\cdot\vA(\vr,t)^*}
\,,
\label{eq-average_order-parameter}
\end{eqnarray}
quantifies the degree of symmetry breaking at time $t$ during the quench.
The maximum magnitude is set by the amplitude of a homogeneous chiral ground state, either
$\vA_+ = \Delta(T)\,e^{i\phi_+}\,\ve_+$ or $\vA_- = \Delta(T)\,e^{i\phi_-}\,\ve_-$. 
However, following a quench finite-size domains of both chiralities are generated, in addition to vortices, anti-vortices and domain walls, and neither type of domain is homogeneous. 

In a typical temperature quench, the order parameter, $\langle\Delta\rangle_t$, starts from the floor of thermal noise well above $T_c$, remains small during an impulse period after the temperature passes below $T_c$, then rapidly grows in magnitude after the patches of nearly uniform order parameter phase become sufficiently large. 
The onset of this rapid growth coincides with the Kibble-Zurek freeze out time $\hat{t}$ when order parameter patches become causally connected.
The order parameter dynamics depends on the quench rate $1/\tau_Q$ and the damping parameter $\gamma$. 
We carried out numerical solutions of the stochastic field equations for a range of parameters: damping and noise from the weak to overdamped regimes, 
$\gamma\in\{ 1.0, 0.5, 0.1, 0.05, 0.01, 0.005 \}$,  
and
$\tau_Q/\btauGL\in$ $\{70,$ $100, 150, 200, 250, 300,400, 500, 750, 1000, 2000, 5000 \}$.

The top row of Fig.~\ref{fig-freeze-out} highlights how the order parameter dynamics changes dramatically as a function of damping after freeze out.
As the damping parameter decreases the order parameter $\langle\Delta\rangle_t$ develops substantial oscillations that decay slowly in time.
These order parameter oscillations are signatures of weakly damped propagating Higgs and clapping modes in the underdamped regime.  

We note that in the literature there are different methods of identifying the freeze-out time (or ``equilibration time'').~\cite{glu25,ant99,suz25,shi26,kir25}
The KZ freeze-out time $\hat{t}$ is identified in our numerical simulations by the time when $\langle \Delta \rangle_t$ reaches $0.01 \kb T_c$, five times larger than the noise floor at $\sim 0.002 \kb T_c$, but 
sufficiently small that coarsening due to nonlinearities is insignificant.
We have verified that the Kibble-Zurek scaling exponents are insensitive to the threshold up to $0.1\,\kb T_c$.
The freeze-out time is evaluated for a wide range of quench times, $\tau_Q$, as shown in the upper row of Fig.~\ref{fig-freeze-out} for the overdamped ($\gamma=1.0$) and underdamped ($\gamma=0.01$) regimes.
We find that this approach captures the correlation length scaling most accurately before coarsening has become significant, and yields scaling exponents in close agreement with scaling theory for strong and weak damping.
In particular, based on the mean field exponent $\nu=1/2$ for the GL correlation length the KZ freeze out time is expected to scale with the quench time as 
\begin{equation}
\hat{t}
\propto\tau_Q^{\deltaKZM}\,,\quad\mbox{where}\quad\deltaKZM = \frac{z\nu}{z\nu+1}
\,,
\end{equation}
which is $\deltaKZM=1/2$ in the overdamped limit ($z=2$),
and $\deltaKZM=1/3$ in the underdamped limit ($z=1$).~\cite{lag97,suz25}
See Ref.~\onlinecite{glu25} for a short review of the KZ scaling relations.

The KZ scaling exponent, $\deltaKZM$, for each value of the damping is obtained from the nonlinear least-squares fits to the freeze-out time shown in the lower left panel of Fig.~\ref{fig-freeze-out}.
The corresponding numerical results for the KZ scaling exponent as a function of $\gamma$ are shown in bottom right panel of Fig.~\ref{fig-freeze-out}, exhibiting a smooth crossover from the underdamped ($\mbox{UD}$) to the overdamped ($\mbox{OD}$) regime. In particular, $\deltaKZM$ is well approximated by
\begin{equation}
\deltaKZM=\deltaUD\,e^{-b \gamma}+\deltaOD\,(1-e^{-b\gamma})
\,,
\end{equation}
with parameters $b=7.43\pm 0.63$, $\deltaUD=0.313 \pm 0.004$ and $\deltaOD=0.488 \pm 0.004$.
These limiting values for the scaling exponents are slightly suppressed relative to the mean-field values for the overdamped ($\deltaKZM=1/2$) and underdamped regimes ($\deltaKZM = 1/3$). The deviations may vary depending on the precise condition used to determine the freeze-out time from numerical simulation data.

\vspace*{-5mm}
\subsection{Collective Mode Oscillations during a Quench}\label{sec-modes_via_quench}
\vspace*{-3mm}

Following a temperature quench, chiral domains form during the inflationary period after freeze-out.
The order parameter dynamics during this time period involves gradients of the order parameter, including the condensate momentum, $\vp_s=\nicefrac{\hbar}{2}\mathbf\nabla\phi$, that source collective modes of the chiral domains.~\cite{sau15}

Here we describe results for a representative simulation with $\tau_Q=500\btauGL$ and $\gamma=0.01$ that reveal the collective modes generated by a quench.
We evaluate $\langle\Delta\rangle_t$ defined by Eq.~\eqref{eq-average_order-parameter}, and similarly $\langle E^\pm\rangle_t$, for a quench in which the multi-domain order parameter that develops just after freeze out evolves to a dominant chiral domain, $A_+(\vr,t)$, with space-time fluctuations relative to the equilibrium amplitude $\Delta$, and a relatively small sub-dominant order parameter, $A_-(\vr,t)$, contributing to the fluctuations.

In the left panel of Fig.~\ref{fig-onset_of_modes}, $\langle\Delta\rangle_t$ is shown in comparison with a reference equilibrium order parameter at the instantaneous temperature, $\Delta(T(t))$.
Substantial oscillations develop at the end of the inflationary period and decay as the temperature approaches the target temperature at $t=\tau_Q/2=250\,\btauGL$.
In the middle panel large oscillations of $\langle E^\pm\rangle_t$ onset early during the period in which $A_+$ and $A_-$ domains are competing and there are substantial gradients exciting the clapping modes.
The oscillations continue long after reaching the target temperature at $\tau_Q/2$. 

Early coarsening dynamics and domain annealing take place during the early post-freeze-out period leading to a dominant $A_+$ domain (in this example) and a subdominant $A_-$ order parameter contributing to fluctuations of the inhomogeneous chiral domain.
In this simulation, the last $A_-$ domain is eliminated at $t \approx 215 \bar{\tau}_\textrm{GL}$.
Fourier transform of $\langle E^+ \rangle_t\rightarrow \langle E^+ \rangle_\omega$ over the time window $t\in [270,800]\,\btauGL$, shown in the right panel of Fig.~\ref{fig-onset_of_modes}, shows a sharp spectral signature of the clapping mode at the expected mass of $\sqrt{2}\Delta$.
Also shown is the Fourier transform of the spatially averaged amplitude, $\langle\Delta\rangle_\omega$, which shows the spectral signature of the Higgs mode at $2\Delta$. This is expected for a single $A_+$ domain since in this case 
$\Delta(\vr,t)\approx\Delta+D^{+}(\vr,t)
+
\cO\left[(D^{+})^2, (D^{-})^2, (E^{+})^2, (E^{-})^2 \right]$.
A spectral signature of the clapping modes at $2\times M_{\text{E}^{\pm}}=2\sqrt{2}\Delta$ also appears as a result of the quadratic contribution of $E^{\pm}$ to $\Delta(\vr,t)$.
The peak in spectral weight at $\omega = 0$ in $\langle\Delta\rangle_\omega$ is expected from the quadratic contribution of $(D^-)^2$, i.e. the Nambu-Goldstone mode. 
However, slow drifting of the background order parameter toward equilibrium as shown in the inset of the left panel of Fig.~\ref{fig-onset_of_modes} may also contribute to the spectral weight near $\omega=0$.
In this simulation we were not able to differentiate these two contributions. The simulation also does not resolve a spectral signature at $2\times M_{\text{D}^{+}}=4\Delta$ from the $(D^{+})^2$ contribution to $\Delta(\vr,t)$.
In Sec.~\ref{sec-nonlinear}, we compute the spectral features for large amplitude fluctuations, specifically the spectral features of the clapping and phase modes 
generated by a large Higgs excitation in comparison to a weak Higgs excitation.

For a quench of a large scale system in which multiple chiral domains remain in the long time limit the spectral signatures of the collective modes are the same; the Higgs mode for an $\ell_z =+1$ or $\ell_z = -1$ domain has mass $2\Delta$, and the corresponding clapping modes for either chiral domain are at $\sqrt{2}\Delta$.
We anticipate that the excitation spectrum will also exhibit spectral signatures due to domain wall motion as well as vortex/anti-vortex annihilation processes. 
Exploring these contributions to the spectrum is beyond the scope of this work.

\vspace*{-5mm}
\subsection{Nonequilibrium dynamics and scaling in $\vk$ space}
\vspace*{-3mm}

The spectrum of excitations based on Fourier amplitudes of the inhomogeneous order parameter provides additional insight into signatures of collective excitations and defects generated by a temperature quench.
The authors of Ref.~\onlinecite{hin00} argue that Fourier amplitudes of the order parameter fall out of equilibrium when
\begin{equation}
\left|\frac{d\omega(k)}{dt}\right| > \left| \omega(k) \right|^2
\,.
\end{equation}
As a result order parameter modes fall out of equilibrium for wavevectors below a critical value, $\hat{k}$, 
that scales as 
$\hat{k}\propto\tau_Q^{-1/4}$ in the overdamped limit, and 
$\hat{k}\propto\tau_Q^{-1/3}$ in the underdamped limit.

For the chiral p-wave superfluid, under conditions of a temperature quench in the weak damping limit, we calculate $|A_+(\vk,t)|^2$, averaged over the ensemble of simulations of the stochastic TDGL equations, and all directions $\varphi_\vk$ of the wave vector, 
\begin{equation}
S(k)=\frac{1}{L^2}\Big\langle\int_0^{2\pi}\frac{d\varphi_\vk}{2\pi}|A_+(\vk)|^2\Big\rangle_{\Omega}
\,.
\label{eq-nonequilibrium_modes-Sk}
\end{equation}

\begin{figure}
\centering
\includegraphics[width=0.9\columnwidth]{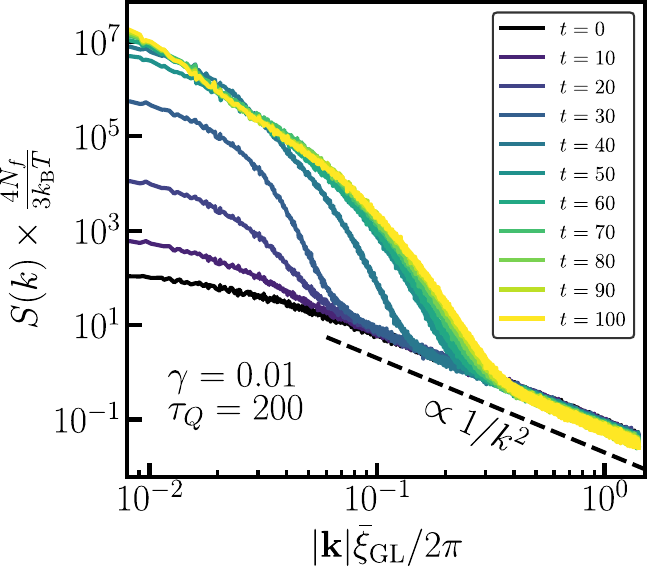}
\caption{
Nonequilibrium excitations of the order parameter $S(k,t)$ vs. $k$ in the weakly damped regime with $\gamma = 0.01$ and a quench time of $\tau_Q/\btauGL = 200$ are obtained from simulations of the stochastic TDGL equations and Eq.~\eqref{eq-nonequilibrium_modes-Sk} as described in the text. 
Each curve corresponds to a time $t$ in units of $\btauGL$ after the temperature passes through $T_c$ at $t=0$.
The longer wavelength modes of the order parameter grow and coarsen as a function of time. 
At each time slice, there is a maximum wavevector, $k_{\text{max}}(t)$, above which $S(k,t)$ scales $\propto 1/k^2$ characteristic of excitations in thermal equilibrium. 
Nonequilibrium excitations below $k_{\text{max}}(t)$ are identified by the enhanced density above the $\propto 1/k^2$ floor.
}
\label{fig-noneq_modes_vs_k}
\end{figure}

We carried out 20 simulations on a system of size $L\times L=440\,\bxiGL\times 440\,\bxiGL$, with $N_k \times N_k = 880 \times 880$ \vk-space grid points. The larger system size allows us to capture the long-wavelength mode dependence and minimize finite size effects.
At large $k$, $S(k)$ scales as $\propto 1/k^2$ characteristic of thermal equilibrium. 
Nonequilibrium excitations appear as deviations in the amplitude above the $1/k^2$ floor for  wavevectors $k\le k_{\text{max}}$, where $k_{\text{max}}$ is defined as the maximum wave number excited above equilibrium and increases with time after the bath temperature passes through $T_c$, as shown in Fig.~\ref{fig-noneq_modes_vs_k}.

%---------------------------------------------------------------------
%---------------------------------------------------------------------

At the KZ freeze-out time we define $\hat{k}\equiv k_{\text{max}}(\hat{t})$, and check for scaling with the quench time, $\hat{k}\propto \tau_Q^{-\kappaKZM}$.
One is tempted to identify $\hat{k}=2\pi/\hat{\xi}$, where $\hat\xi\equiv\xi(T(\hat{t}))\propto\tau_Q^{\chiKZM}$ is the correlation length at freeze-out, in which case we would also expect 
$\kappa_{\textrm{KZM}}=\chiKZM$.
The mean-field prediction for the scaling exponent is then $\chiKZM=\frac{\nu}{z\nu+1}=\kappaKZM$. 
The scaling exponent $\kappa_{\textrm{KZM}}$ depends on the damping rate, $\gamma$, 
but in contrast to the correlation length and freeze-out time, 
the scaling relation for $\hat{k}$ is difficult to determine as it is sensitive to the criteria adopted to identify KZ freeze-out.
We test the scaling relation for the correlation length in real space in the next section.

\vspace*{-5mm}
\subsection{Spatial correlations of the order parameter} 
\vspace*{-3mm}

From the $\vk$-space spectrum of excitations we can obtain the spatial correlation function for the order parameter from the Wiener-Khinchin theorem, i.e. the $\vk$-space mode amplitude and correlation function $C(\vr,t)\ns=\ns\langle A_+(\textbf{x},t) A^*_+(\textbf{x}+\vr,t) \rangle_{\vx,\Omega}$ are related by
\begin{equation}
\hspace*{-2mm}
C(\vr,t)=\left\langle \int\ns\dfrac{d^2k}{(2\pi)^2}\,|A_+(\vk,t)|^2 e^{i\vk\cdot\vr} \right\rangle_{\Omega}
\,.
\end{equation}
The result is then averaged over all angles $\varphi_\textbf{r}$ to generate $C(r)$.
In the homogeneous ground state the superfluid has long-range order with 
$\lim\limits_{r\rightarrow\infty} \frac{C(r)}{C(0)}=1$.
However, at any finite time following a temperature quench, the inhomogeneous condensate is far from homogeneous equilibrium, and as a result $C(r,t)$ decays toward zero on a length scale which depends both on the quench rate and the measurement time.
Over longer timescales the inhomogeneous condensate coarsens and $C(r)$ grows.

%---------------------------------------------------------------------
\begin{figure}[t!]
\centering
\includegraphics[width=\columnwidth]{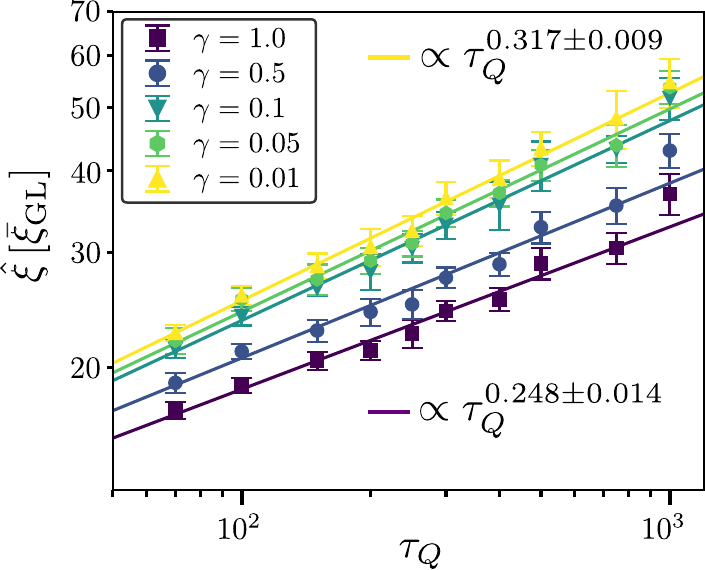}
\caption{
The correlation length at KZ freeze-out, $\hat\xi$, obtained from the order parameter correlation function $C(r,t)$ evaluated at $t=\hat{t}$ as a function of the quench time, $\tau_Q$, and damping parameter, $\gamma$.
The log-log plot shows the scaling relation, $\hat\xi\propto \tau_Q^{\chiKZM}$, and yields the exponent, $\chiKZM$, as a function of $\gamma$, with limiting values close to the mean-field exponents of $\chiKZM=1/3$ for $\Gamma\rightarrow 0$ and $\chiKZM=1/4$ for $\mu\rightarrow 0$.
}
\label{fig-correlation_length_freeze-out}
\end{figure}
%---------------------------------------------------------------------

We identify the correlation length at the KZ freeze-out time, $\hat{\xi}$, with the half-width at half-maximum 
of the correlation function, $C(\hat{\xi},\hat{t})=0.5 C(0,\hat{t})$.
The mean values of $\hat{\xi}$ are computed by using the ensemble averaged spatial correlation functions, while the error bars are the standard deviation of $\hat{\xi}$ computed over individual simulations.
The resulting analysis of our simulations yields $\hat{\xi}$ as a function of the quench time, $\tau_Q \in \{70, 100, 150, 200, 250, 300, 400, 500, 750, 1000\}\btauGL$, and the damping parameter, $\gamma$ (in this calculation, $\gamma \in \{1.0, 0.5, 0.1, 0.05, 0.01\}$), as shown in Fig.~\ref{fig-correlation_length_freeze-out}.
Thus, the KZ scaling exponents for the correlation length at freeze-out vary from $\chiKZM=0.317\pm 0.009$ in the underdamped regime ($\gamma=0.01$) to $\chiKZM=0.248\pm 0.014$ in the overdamped regime ($\gamma=1.0$), and are in reasonable agreement with the mean-field predictions of $\chiKZM=1/3$ and $\chiKZM=1/4$, respectively.

\begin{figure}
\centering
\includegraphics[width=\columnwidth]{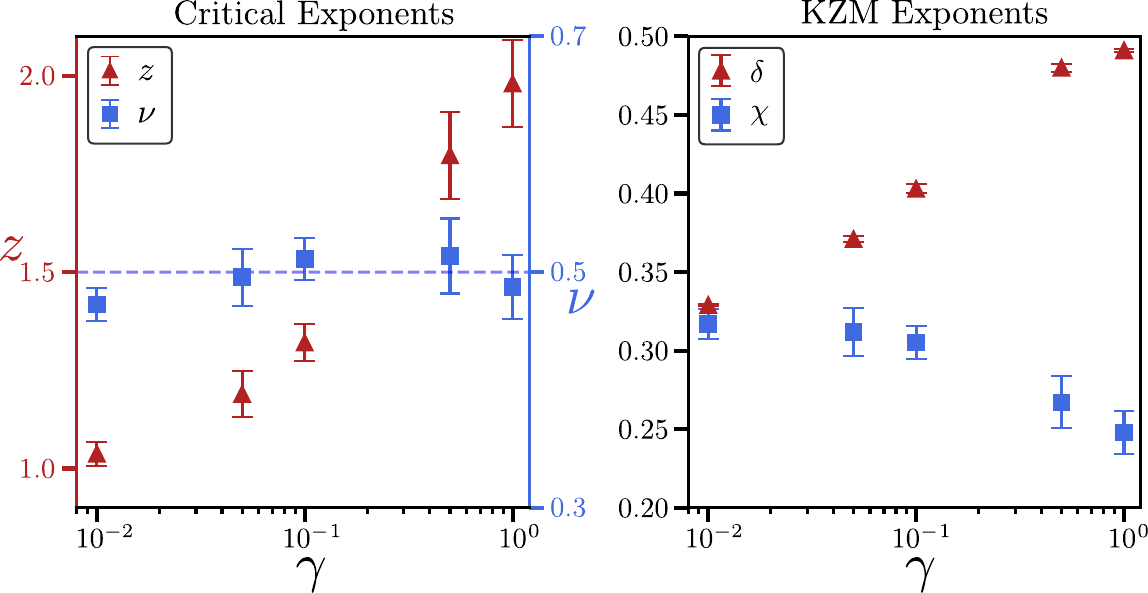}
\caption{
Left panel: scaling of dynamic exponent $z$ and equilibrium correlation length exponent $\nu$ as a function of damping.
The dynamic exponent varies smoothly from 1 to 2 as damping is increased, while $\nu \approx 1/2$ is independent of damping (see dashed blue line).
Right panel: scaling of Kibble-Zurek exponents with damping.
In the low-damping regime, $\deltaKZM = \chiKZM = 1/3$.
As damping increases, $\deltaKZM$ increases to $\sim 1/2$ while $\chiKZM$ decreases to $\sim 1/4$.
}
\label{fig-exponents}
\end{figure}

Our numerical simulation results for the correlation length exponent, $\nu$, and dynamic exponent, $z$, are shown in Fig.~\ref{fig-exponents}, which are extracted from $\delta_{\textrm{KZM}} = z\nu / (z\nu + 1)$ and $\chiKZM = \nu / (z\nu + 1)$.
Notable is that $\nu$ is independent of damping, while $z$ varies smoothly with damping from $z\approx 1$ in the underdamped regime to $z\approx 2$ in the overdamped regime.

\vspace*{-5mm}
\section{Nonlinearities and the mode spectrum}\label{sec-nonlinear}
\vspace*{-3mm}

The nonlinear terms in the TDGL field equations for the complex vector order parameter, $\vA(\vr,t)$, can be expressed in terms of four excitation amplitudes for chirality $\ell_z=+1$, $D^{\pm}(\vr,t)$, and for chirality $\ell_z=-1$, $E^{\pm}(\vr,t)$. 
In thermal equilibrium below $T_c$, or during the initial time period of a temperature quench, the order parameter 
fluctuations are sufficiently small that the nonlinear terms can to first approximation be dropped. 
In this limit the four amplitudes decouple and correspond to the amplitudes for the spectrum of collective modes of a homogeneous region of order parameter with well defined masses as described in Sec.~\ref{sec-OP_modes}. 
Thus, the masses of the Higgs and clapping modes are well defined in the equilibrium PSD for weak damping shown in Fig.~\ref{fig-equilibrium_mode_PSD}.

%---------------------------------------------------------------------
%---------------------------------------------------------------------

However, as the order parameter amplitude increases rapidly post freeze-out the nonlinear terms in the stochastic TDGL equations lead to (1) coupling with the Higgs mode at quadratic order, and (2) third-harmonic generation via cubic terms, e.g. $D^+\,(E^+)^2$, $E^{\pm}(D^+)^2$ etc.
The dominant nonlinearity is always quadratic with at least one $D^+$ term.

These nonlinearities significantly modify the PSD obtained for thermal equilibrium for the time period near freeze-out. 
The presence of finite amplitude fluctuations leads to harmonic generation, mass shifts, and the generation of additional resonances in the PSD.

\begin{figure}
\centering
\includegraphics[width=\linewidth]{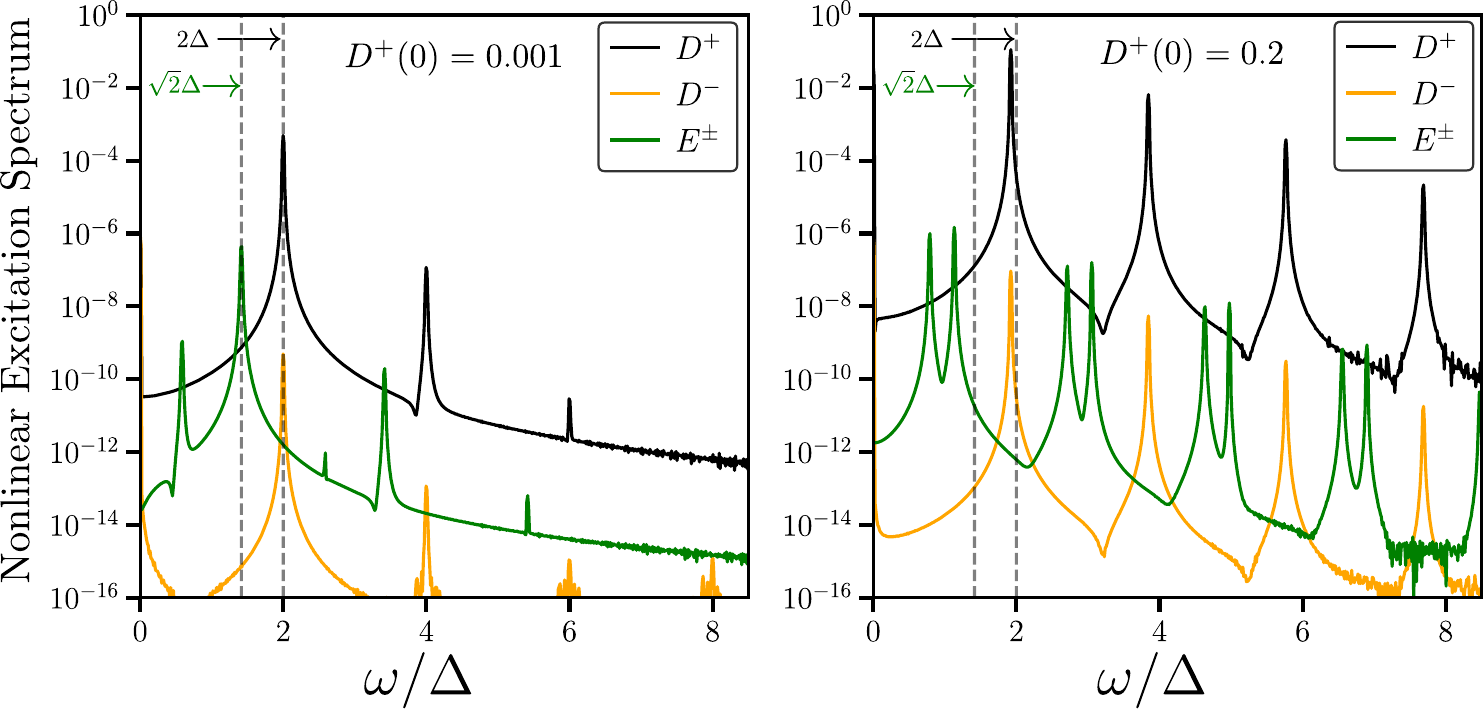}
\caption{
Fourier transform of excitation amplitudes for numerical solutions to the nonlinear homogeneous TDGL equations with initial conditions of a small Higgs amplitude $D^+=0.001\Delta$ (left panel) and a large Higgs amplitude, $D^+=0.2\Delta$ (right panel). 
Dashed vertical lines at $\sqrt{2}\Delta$ and $2\Delta$ correspond to the masses of the clapping and Higgs modes. 
For small amplitude excitations there are resonant peaks at $\omega=2\Delta$ (Higgs mode) and $\omega=\sqrt{2}\Delta$ (clapping modes), as well as weak harmonic generation and a $2\Delta$ resonance in the phase mode from the quadratic nonlinearity.
For a large initial nonequilibrium Higgs amplitude, $D^+=0.2\Delta$, we obtain large mass corrections, and the generation of multiple resonances in the spectrum.
}
\label{fig:homogeneous_nonlinear}
\end{figure}

The presence of defects does not allow the amplitudes $D^\pm$, $E^\pm$ to be a sufficient description of the whole system, given there are multiple background superfluids separated by domain walls.
Here we consider the effect of nonlinearities, generated from the $3^{rd}$ line of the TDGL Lagrangian in Eq.~\eqref{eq-Lagrangian}, in a homogeneous system with zero damping.
In Fig.~ \ref{fig:homogeneous_nonlinear} we show results for the Fourier transform of the order parameter for two different initial conditions for the Higgs amplitude. 
In the left panel we set $D^+(0)=0.001\Delta$, and all other excitation amplitudes to $10^{-6}\Delta$.
The initial time derivatives of all excitation amplitudes are set to zero, and the equations of motion are solved with Verlet integration and a time step of $\Delta t = 10^{-4}\btauGL$ for a duration of 500 $\btauGL$.
The resulting signal is multiplied by a Hann window function and the Fourier transform of the solution shows the Higgs excitation at $M=2\Delta$ and the clapping modes at $M=\sqrt{2}\Delta$. 
The quadratic nonlinearity also contributes a $2\Delta$ resonance to the massless $D^-$ phase mode.
For the results shown in the right panel we initialized the equations with a larger Higgs amplitude,
$D^+(0) = 0.2 \Delta$, relevant to large oscillations in a quench, but kept all other initial conditions the same.
The nonlinear terms generate a complex spectrum resulting from harmonic generation and mode-mode coupling.
Note that the results in Fig.~\ref{fig:homogeneous_nonlinear} are for the Fourier transform, and squaring the results gives the PSD.

%---------------------------------------------------------------------
%---------------------------------------------------------------------

%---------------------------------------------------------------------
\vspace*{-5mm}
\section{Conclusion}
\vspace*{-3mm}

\ul{Summary:} The ground state of \He\ films is the chiral p-wave A-phase which spontaneously breaks time-reversal and mirror reflection symmetry.
The ground state is doubly degenerate with the chiral axis parallel or anti-parallel to the normal to the 
film surface, i.e. $\hat{\ns\vell}||\pm \hat\vz$. 
As a result \Hea\ films host both quantized vortices and domain walls separating degenerate chiral domains.
Our numerical simulations of the nonequilibrium phase transition support the Kibble-Zurek mechanism for topological defect generation, with the freeze-out time and correlation length obeying KZ scaling relations.

This study highlights the order parameter dynamics and generation of complex inhomogeneous ordered condensates based on stochastic TDGL field equations, with an emphasis on the generation of bosonic collective excitations of the inhomogeneous chiral condensate by a nonequilibrium temperature quench into the superfluid phase. 
Langevin noise and damping, coupled to the space-time dynamics of the order parameter, are shown to generate a power spectrum of bosonic excitations that are signatures of the broken symmetry chiral ground state.
In a temperature quench to $T < T_c$ bosonic excitations are generated immediately following the inflationary period post Kibble-Zurek freeze-out when the bosonic PSD is enhanced and modified by nonlinearities in the stochastic TDGL field equations. 
The order parameter correlation function for bosonic excitations exhibits scaling relations with exponents for the real space correlation length at freeze out, and highlights the smooth cross-over in the dynamic exponent as a function of the damping rate.

\ul{Outlook:} Nonequilibrium space-time dynamics of multi-component topological superfluids is a frontier for the study of nonequilibrium phase transitions, the generation and dynamics of topological defects and the spectrum of collective excitations. Superfluid \He, which is an ultra-pure quantum fluid,  provides a unique laboratory system for the study of metastability and nonequilibrium phase transitions widely studied theoretically in the context of phase transitions in early universe cosmology, as well as cold dense nuclear matter in neutron stars.

%---------------------------------------------------------------------
\vspace*{-5mm}
\section{Acknowledgements}
\vspace*{-3mm}

We thank Mark Hindmarsh for valuable discussions on computational field theory simulations. 
This research was supported by the Hearne Institute of Theoretical Physics and the 
Center for Computation and Technology at Louisiana State University.

%---------------------------------------------------------------------
\appendix

\vspace*{-5mm}
\section{TDGL material parameters}\label{app-TDGL_parameters} 
\vspace*{-3mm}

The coefficient $\mu$ determines the order parameter inertia and kinetic energy term in the TDGL 
Lagrangian.
Fixing $\mu$ by the mass of the Higgs mode, $2\Delta$, leads to different inertial coefficients in the A- and B-phases of \He.~\cite{sau15,sau17} In the case of \Hea,
\begin{equation}
\mu=\frac{7\zeta(3)}{30} N_f \left(\frac{\hbar}{2\pi \kb T_c} \right)^2 = \beta_{245}\hbar^2
\label{eq-mu}
\,.
\end{equation}
This coefficient also defines a time scale for inertial order parameter dynamics,  
\begin{equation}
\bar\tau_{\text{GL}}
= \sqrt{\frac{7\zeta(3)}{30}}\frac{\hbar}{2\pi \kb T_c}
= \sqrt{\frac{7\zeta(3)}{30}}\tau_0
\,.
\label{eq-tauGL}
\end{equation}
The term $\alpha(T)$ defines the onset of condensation and superfluidity at the transition 
temperature, $T_c$, and varies linearly in temperature, changing sign at $T_c$,
\begin{equation}
\alpha(T) = \frac{N_f}{3}\left( \frac{T}{T_c} - 1 \right)
\,.
\label{eq-alpha}
\end{equation}

The gradient coefficients determine the cost in energy associated with spatial variations of the order parameter, as well as the dispersion of bosonic collective modes in TDGL theory. 
In the weak coupling limit,
\begin{equation}
K_1=K_2=K_3=\frac{7\zeta(3)}{60}N_f\left(\frac{\hbar v_f}{2\pi\kb T_c}\right)^2
\,.
\label{eq-K123}
\end{equation}
The ratio of the gradient energy to the condensation energy also defines the minimum length scale
for spatial variations of the order parameter,
\begin{eqnarray}
\xiGL(T) &=& \sqrt{\frac{K_1}{|\alpha(T)|}} = \frac{\bxiGL}{(1-T/T_c)^{\tinyonehalf}}
\,,
\\
\mbox{where}\quad\bxiGL &=& \sqrt{\frac{7\zeta(3)}{20}}\frac{\hbar v_f}{2\pi \kb T_c}
\,.
\end{eqnarray}
The $\beta$ coefficients for each of the nonlinear terms in the TDGL action are essential for determining the ground states of superfluid \He,~\cite{wim16,reg20} as well as the excitation energies of the bosonic modes of \He.~\cite{sau17,wim18}
In the weak-coupling approximation the five $\beta$ coefficients are given by fixed ratios,
\begin{equation}
-2 \betawc_1 = \betawc_2 = \betawc_3 = \betawc_4 = - \betawc_5 = 2\beta_0
\,,
\label{eq-betas_wc}
\end{equation}
\begin{equation}
\mbox{where}\quad\beta_0=\frac{7\zeta(3)}{240}\frac{N_f}{(\pi\kb T_c)^2}
\,.
\label{eq-beta0}
\end{equation}
Leading order strong-coupling corrections to weak-coupling beta parameters, $\betasc_i$, are essential 
for a quantitative theory for the stability of bulk, confined and inhomogeneous phases 
of \He.~\cite{wim16,wim19,reg20}
The resulting pressure and temperature dependent $\beta$ parameters are given by~\cite{wim16}
\begin{equation}
\beta_i(T,p) = \betawc_i(p) + \frac{T}{T_c}\betasc_i(p)\,,\quad i\in\{1,\ldots,5\}
\,.
\label{eq-betas_sc-scaling}
\end{equation}
The most accurate theoretical results for the strong-coupling $\betasc_i$ parameters reproduce the 
pressure-temperature phase diagram,~\cite{wim19} in close agreement with experiment.~\cite{gre86}

In the case of thin \He\ films, which is the focus of this report, the chiral A-phase and non-chiral 
planar phase are degenerate in weak-coupling theory. Strong-coupling corrections lift the degeneracy 
and stabilize the chiral A-phase at all pressures and all temperatures down to $T=0$
where strong-coupling corrections are extremely small.

\vspace*{-7mm}
\section{Mode-mixing terms \&  the equilibrium PSD}\label{app-K23} 
\vspace*{-3mm}

Retaining the quadratic mode-mode coupling terms in the Lagrangian and converting to the chiral basis, we have
\begin{eqnarray}
L=\int_{\text{V}}d^3r\,
\Big\{
\mu\,\Big[(\dot{D}^{+})^2 + (\dot{D}^{-})^2 + (\dot{E}^{+})^2 + (\dot{E}^{-})^2\Big]
\nonumber\\
-4\Delta^2\Big[\bar\beta_1(D^{+})^2 +\bar\beta_2\left((E^{+})^2 + (E^{-})^2
\right)\Big] 
\nonumber\\
-2 K_1\Big[
(\partial_i D^+)^2 + (\partial_i D^-)^2 + (\partial_i E^+)^2 + (\partial_i E^-)^2
\nonumber\\
+ \partial_x D^+ (\partial_x E^+ + \partial_y E^-) + \partial_y D^+(\partial_x E^- - \partial_y E^+)
\qquad
\nonumber\\
+ \partial_x D^- (\partial_x E^- - \partial_y E^+) - \partial_y D^- (\partial_x E^+ + \partial_y E^-)
\Big]
\Big\} 
\,.\,\,\,\,\,\,
\label{eq-K23_Lagrangian}
\end{eqnarray}
The latter two rows of gradient terms mix the excitations at quadratic order.
The coupled dynamical equations after spatial Fourier transform are
\begin{widetext}
\begin{align}
    \ddot{D}^+ + (\Gamma/\mu) \dot{D}^+ + 4 \Delta^2 D^+ + v^2 (k_x^2 + k_y^2) D^+
    +\nicefrac{1}{2}v^2\left(k_x^2 - k_y^2 \right) E^+ + v^2 k_x k_y E^- = \zeta_{D^+} (\vk, t)/\mu\,,
    \nonumber\\
    \ddot{D}^- + (\Gamma/\mu)\dot{D}^- \qquad\qquad + v^2 (k_x^2 + k_y^2) D^-
    +\nicefrac{1}{2}v^2\left(k_x^2 - k_y^2 \right) E^- - v^2 k_x k_y E^+ = \zeta_{D^-} (\vk, t)/\mu\,,
    \nonumber\\
    \ddot{E}^+ + (\Gamma/\mu)\dot{E}^+ + 2 \Delta^2 E^+ + v^2 (k_x^2 + k_y^2) E^+
    +\nicefrac{1}{2}v^2\left(k_x^2 - k_y^2 \right)D^+ - v^2 k_x k_y D^- = \zeta_{E^+} (\vk, t)/\mu\,,
    \nonumber\\
    \ddot{E}^- + (\Gamma/\mu)\dot{E}^- + 2 \Delta^2 E^- + v^2 (k_x^2 + k_y^2) E^-
    +\nicefrac{1}{2}v^2\left(k_x^2 - k_y^2 \right)D^- + v^2 k_x k_y D^+ = \zeta_{E^-} (\vk, t)/\mu\,.
\end{align}
\end{widetext}
Defining $s = \frac{1}{2}v^2 |\vk|^2$ and $\theta = \tan^{-1}(k_y / k_x)$, the gradient terms become $v^2(k_x^2 + k_y^2) = 2s$, $\frac{v^2}{2}(k_x^2 - k_y^2) = s\cos(2\theta)$, $v^2 k_x k_y = s\sin(2\theta)$.
The equations can be cast into a matrix form as
\begin{equation}
\ddot{\textbf{X}} + \frac{\Gamma}{\mu} \dot{\textbf{X}} + \vM\,\textbf{X}= \boldmath{\zeta}/\mu,
\end{equation}
where the ``mass matrix'' is
\begin{equation}
\textbf{M} = \begin{pmatrix}
4\Delta^2 + 2 s & 0 & s \cos(2\theta) & s \sin(2 \theta) \\
0 & 2 s & -s \sin(2\theta) & s\cos(2\theta) \\
s \cos(2\theta) & -s \sin(2\theta) & 2\Delta^2 + 2 s & 0 \\
s \sin(2\theta) & s \cos(2\theta) & 0 & 2\Delta^2 + 2 s
\end{pmatrix}
\end{equation}
Here the $2\times 2$ off-diagonal blocks are $s$ times a rotation matrix of angle $2\theta$.
The mode coupling can be simplified by applying a rotation on the $E$ variables:
\begin{equation}
\begin{pmatrix}
E^{+\prime} \\ E^{-\prime}
\end{pmatrix}
=
\begin{pmatrix}
\cos(2\theta) & \sin(2\theta) \\
-\sin(2\theta) & \cos(2\theta)
\end{pmatrix}
\begin{pmatrix}
E^{+} \\ E^{-}
\end{pmatrix}
\end{equation}
After the transformation, the matrix in the new basis is
\begin{equation}
\textbf{M}^{\prime} 
=
\begin{pmatrix}
4\Delta^2 + 2 s & 0 & s & 0 \\
0 & 2 s & 0 & s \\
s & 0 & 2\Delta^2 + 2 s & 0 \\
0 & s & 0 & 2\Delta^2 + 2 s
\end{pmatrix}
\end{equation}
The noise term is independent for each excitation, so the noise is statistically equivalent following the basis transformation.
Now only excitations with the same charge conjugation parity are coupled through the gradient terms, i.e. $D^+$ with $E^{+\prime}$ and $D^-$ with $E^{-\prime}$.
Thus, the $4\times 4$ matrix equations reduce to two subsystems of $2\times 2$ matrices, 
\begin{equation}
\ddot{\vX}^c + \frac{\Gamma}{\mu} \dot{\textbf{X}}^c + \vm^c \textbf{X}^c = \boldmath{\zeta}^c /\mu,
\end{equation}
where the superscript $c$ labels the charge conjugation parity, and the $2\times 2$ sub-matrices are,
\begin{equation}
\hspace*{-2.5mm}
\vm^+\ns=\ns\begin{pmatrix} 4\Delta^2\ns+\ns 2s & s \cr s & 2\Delta^2\ns+\ns 2s\end{pmatrix},\,
\vm^-\ns=\ns \begin{pmatrix} 2s & s \cr s & 2\Delta^2\ns+\ns 2s\end{pmatrix}.
\end{equation}
Fourier transforming to the frequency domain gives,
\begin{equation}
(\vm^c - \omega^2 - i\Gamma\omega/\mu)\vX^c = \boldsymbol{\zeta}^c/\mu\,.
\end{equation}
Defining the matrix operator on the left side of the equation as 
$[\vG^c]^{-1}(s,\omega)=\vm^c-\omega^2-i\Gamma\omega/\mu$, allows us to express 
the momentum dependent PSD for a given collective mode $i$ with parity $c$ as
\begin{equation}
\mathcal{P}^c_i(s,\omega)=\frac{\Gamma \kb T}{\mu^2} \sum\limits_j \left|G^c_{ij}(s,\omega) \right|^2
\label{eq-PSD_vs_s}
\end{equation}
Straightforward matrix identities, 
$2i\,\Im\vG=\vG-\vG^\dagger=\vG[(\vG^\dagger)^{-1}-\vG^{-1}]\vG^\dagger
=
\frac{2i\,\Gamma\omega}{\mu}\vG\vG^\dagger$, 
for either $c$-parity allow us to express the PSD as
\begin{equation}
\mathcal{P}^c_i(s,\omega)=\frac{\kb T}{\mu \omega} \Im\left[G^c_{ii}(s,\omega)\right]
\,.
\label{eq-PSD_vs_s-omega}
\end{equation}
Inverting $[\vG^c]^{-1}$ and noting that $\det\left([\vG^\pm]^{-1}\right)$ are quadratic polynomials in $s$,

\begin{eqnarray}
\det\left([\vG^+]^{-1}\right) = a^+ s^2 + b^+ s + d^+ 
\,,
\label{eq-detG-plus}
\\
\det\left([\vG^-]^{-1}\right) = a^- s^2 + b^- s + d^-
\,,
\label{eq-detG-minus}
\end{eqnarray}
with 
$a^+ = a^- = 3$,
$b^+ = 12\Delta^2 - 4(\omega^2 + i\Gamma\omega/\mu)$, 
$b^- = 4\Delta^2 - 4(\omega^2 + i\Gamma\omega/\mu)$, 
$d^+ = (4\Delta^2 - \omega^2 - i\Gamma\omega/\mu)(2\Delta^2 - \omega^2 - i\Gamma\omega/\mu)$, and 
$d^- = -(\omega^2 + i\Gamma\omega/\mu) (2\Delta^2 - \omega^2 - i\Gamma\omega/\mu)$.
This allows the PSD to be expressed as an integration over wavenumber, or $s=\frac{1}{2}v^2|\vk|^2$, as
\begin{equation}
\mathcal{P}_{X}(\omega) = \frac{\kb T}{2 \pi K \omega} 
\Im\int\limits_0^\infty ds\,
\frac{2s + M_{\bar{X}}^2 - \omega^2 - i\Gamma\omega/\mu}{a_c s^2 + b_c s + d_c}\
\,,
\label{eq-powerX_with_mixing}
\end{equation}
where $K=\mu v^2$ and $M_{\bar{X}}$ refers to the mass of the partner excitation, e.g. 
$D^+$ mixes with $E^{+\prime}$ so $M_{\bar{X}} = \sqrt{2}\Delta$.
We reference each of the four eigenmodes at finite $\vk$ by their $|\vk|=0$ amplitudes and masses, 
$M_{\text{D}^+}=2\Delta$,
$M_{\text{D}^-}=0$,
$M_{\text{E}^{+\prime}}=\sqrt{2}\Delta$ and
$M_{\text{E}^{-\prime}}=\sqrt{2}\Delta$.
For the $D^+$ mode the PSD can be expressed as
\begin{equation}
\mathcal{P}_{D^+}(\omega) = \frac{\kb T}{2\pi K\omega}\,
\Im\int_0^\infty ds\left[\frac{A}{s - s_+} + \frac{B}{s - s_-}\right],
\end{equation}
where $s_\pm$ are the roots of the denominator in Eq.~\eqref{eq-powerX_with_mixing},
$A = (M_{\text{E}^{+\prime}}^2 + 2s_+ - \omega^2 -i\Gamma\omega/\mu)/(a(s_+ - s_-))$, 
$B = (M_{\text{E}^{+\prime}}^2 + 2s_- - \omega^2 -i\Gamma\omega/\mu)/(a(s_- - s_+))$,
and $A+B=2/a$ is real.
We introduce a short wavelength cutoff, $\Lambda$, to regulate the log-divergence of the real part of the integral, then evaluate the imaginary part in the limit $\Lambda\to\infty$ to obtain the cutoff independent result for the PSD,
\begin{eqnarray}
\hspace*{-3mm}\mathcal{P}_{D^+}(\omega) 
\ns&=&\ns 
\lim_{\Lambda\to\infty}\frac{\kb T}{2\pi K \omega}
\Im\Big\{ A\left[\log(\Lambda-s_+) - \log(-s_+) \right] 
\nonumber\\
\ns&+&\ns 
B\left[\log(\Lambda - s_-) - \log(-s_-) \right] \Big\} 
\nonumber\\
\ns&=&\ns 
\frac{-\kb T}{2\pi K \omega} \Im\Big\{ A\log(-s_+) + B\log(-s_-) \Big\}.
\hspace*{3mm}
\label{eq-PSD_final}
\end{eqnarray}
%
%The regulator drops out from $\Im[(A + B)\log \Lambda] = 0$ and $\log(1 - s_\pm / \Lambda) \to 0$.

The PSD for the $E^\pm$ modes are obtained from the PSD for the $E^{+\prime}$ and $E^{-\prime}$ modes. 
In particular for fixed wave number,
\begin{equation}
\mathcal{P}_{E^+}(s,\omega)
=
\cos^2(2\theta)\mathcal{P}_{E^{+\prime}}(s,\omega)
+
\sin^2(2\theta) \mathcal{P}_{E^{-\prime}}(s,\omega)
\,,
\end{equation}
and similarly for $E^-$.
Thus, the angular averaged PSD for the $E^{\pm}$ modes reduces to
\begin{equation}
\hspace*{-3mm}
\mathcal{P}_{E^+}(\omega)
=
\mathcal{P}_{E^-}(\omega)
=\nicefrac{1}{2}\Big[\mathcal{P}_{E^{+\prime}}(\omega)+\mathcal{P}_{E^{-\prime}}(\omega) \Big]
\,,
\end{equation}
where the PSDs on the right-hand side are evaluated similarly to the $D^{\pm}$ modes as described above.

Lastly, integrating Eq.~\eqref{eq-PSD_vs_s-omega} over all frequencies yields the thermal equilibrium mode population per unit area, including the contributions from mode mixing,
\begin{equation}
\frac{\langle |X(\vk)|^2 \rangle}{L^2} 
=
\frac{\kb T}{2 \mu}\,\Re\left[G_{ii}^c (s,\omega=0)\right]
\,,
\label{eq-thermal_population_X} 
\end{equation}
where $c$ and $i$ define mode $X$, and the resulting integration is obtained via the Kramers-Kronig relation,
\begin{equation}
\Re\left[G_{ii}^c (s,\omega=0)\right] 
= 
\dfrac{1}{\pi}\,\textrm{P}\ns\int d\omega^\prime \dfrac{\Im\left[G_{ii}^c (s, \omega^\prime) \right]}{\omega^\prime}.
\end{equation}
Equations~\eqref{eq-thermal_population_X} for the thermal populations of collective modes X are plotted in the left panels of Fig.~\ref{fig-equilibrium_mode_PSD}.
For the clapping modes, the transformation from the $E^{\pm \prime}$ to the physical $E^\pm$ variables is identical to the PSD calculation.

%-------------------------------------------------------
%\bibliography{QFS,CM,SM,Astro,Books,Numerics}
%\bibliographystyle{apsrev4-1-PRXstyle}
%-------------------------------------------------------
%
%
\end{document}